\documentclass[aps,twocolumn]{revtex4}
\usepackage{graphicx}
\usepackage{amsmath}
\usepackage{latexsym}
\usepackage{amsmath}
\usepackage{amssymb}
\usepackage{float}
\usepackage{epstopdf}
\usepackage{hyperref}
\hypersetup{urlcolor=red,citecolor=blue}
\usepackage{bm}
\begin{document}
\title{Hybrid super-lattices of graphene and hexagonal boron nitride: \\
Ferromagnetic semiconductor at room  temperature}
\author{Rita Maji and Joydeep Bhattacharjee\\
\small{\textit{School of Physical Sciences}}\\
\small{\textit{National Institute of Science Education and Research}}\\
\small{\textit{HBNI, Jatni - 752050, Odisha, India}}}
\begin{abstract}
Carbon(C) doped hexagonal boron nitride(hBN) has been experimentally reported to be ferromagnetic at room temperature. 
Substitution by C in hBN has been also reported to form islands of graphene. In this work
we derive a mechanistic understanding of ferromagnetism with graphene islands in hBN from first principles 
and mean-field Hubbard model.
We find a general property, that in bipartite lattices where the sublattices differ in on-site energies, as in hBN, 
the ordering between local magnetic moments can be substantial and predominantly  anti-ferromagnetic(AFM)
if they are embedded in the same sublattice, unless dominated by Mott like inter-sublattice
spin separation due to strong localization.
The dominant AFM order is rooted at spin resolved spatial separation of lone pairs of nitrogen(N) and 
back transferred electrons on boron(B) due to Coulomb repulsion thus essentially implying a super-exchange pathway. 
Subsequently we propose a class of ferri-magnetically ordered inter-penetrating super-lattices of 
magnetic graphene islands in hBN, which can be chosen to be a ferromagnetic semiconductor or a half-metal, 
and retain a net non-zero magnetic moment at room temperature.
\end{abstract}
\maketitle
\section{Introduction}
Possibility of ferromagnetism exclusively due to electrons in 2$p$ orbitals, 
particularly in the functionalized three coordinated bipartite networks
\cite{rev_mag_fm_p_orbital,rev_mag_gnr} of boron(B), carbon(C) and nitrogen(N), 
has opened up a new direction in pursuit of magnetic materials, 
which could be lighter and thinner than those made of traditional metals, 
besides having large spin relaxation time due to weaker spin-orbit coupling.
Designing such materials which can also be magnetically as well as structurally stable at room temperature
is thus a key objective in search for new paradigms of nano-fabrication.
Magnetism in half-filled bipartite systems \cite{lieb,fazekas} are primarily sourced at functionalizations due to structural, 
physical or chemical modification, which impact the two sublattices unequally and hinder $\pi$ conjugation thereby,
leading to ferri-magnetic(FeM) order between the sublattices then in general.
Such functionalized bipartite systems would thus sustain a net non-zero magnetic moment,
accompanied by finite density of states(DOS) for the minority spin at Fermi energy, 
leading to metallic phases allowing spin polarized transport.
Many such scenarios of functionalization of graphene(Gr)\cite{RT_fm-gnr-expt,dft_mag_fm_Hy_gr},
particularly in ribbons \cite{phy-func,hm-2009} and finite segments with magnetic edges
\cite{spin-filter,tri-gnf, grflakes-fm, trignf-chain}, have been widely proposed in the last decade or so.

However, presence of finite DOS near Fermi energy due to the inherent semi-metallic nature of graphene,
and also due to spurious edge states, tend to undermine the effects of functionalizations, 
unless precisely cut into ribbons of systematic band gap, which poses its own experimental challenge.
On the other hand hexagonal boron nitride(hBN) being a wide band gap due,
do not interfere with the spin polarized DOS of graphene islands embedded in it, thus allowing greater
scopes of their manipulation as per device requirements.
Indeed, magnetic Gr-hBN hybrids that have been widely proposed 
\cite{spintronic-BCN, fm_c_bntube, qdot-dft, hybrid-gbn-dft, hbn-interface, c-dope-2018, gr-bn-2018}
in last few years as a promising magnetic material made of non-metals, alongside  
several reports of success in robust synthesis\cite{hetero-expt1,hetero-expt2,ebeam,expt-chbn} of such hybrid structures
with arrays of Gr-islands embedded in hBN.
Triangular Gr-islands with zigzag edges, which offer the largest magnetic moment\cite{tri-gnf} for a given island size, 
are expected to form stable motif\cite{Yakob_NL_2011,expt-chbn,zginter_NL_2014} in hBN.
Neighboring magnetic Gr-islands embedded in hBN have been argued to prefer anti-ferromagnetic(AFM)\cite{afm-rt,c-island} 
ordering in proximity, although a less stable ferromagnetic(FM) phase has also been suggested at some specific proximities of islands. 
C doped hBN has been experimentally reported\cite{expt-fm_BN_GR} to be  ferromagnetic at room temperature in recent years.
These results so far have been mostly anticipated mostly heuristically without any comprehensive closure on a microscopic mechanism. 
Thus the true nature of interaction between localized magnetic moments in hBN needed to be thorough established in order 
to systematically understand the observed results, and also to make rational proposals for new Gr-hBN hybrids with enhanced 
stability of the FM phase. 

Thus the primary task that we set for us in this work, is to understand the exact mechanism of mediation of magnetic order 
by hBN between two Gr-islands with non-zero magnetic moment, followed by focus on stabilizing the FM phase,
which led to systematic proposals for half-metal or ferromagnetic semiconductor with Gr-hBN hybrids, 
with their ferromagnetism sustainable at room temperature.
Our study is based on analysis of spin polarized electronic structures calculated from first principles\cite{dft}, 
as well as within the 
mean-field approximation of Hubbard model\cite{fazekas}(MFH) in a tight-binding(TB) framework to establish
our proposed phenomenological model. 
Effective strength of exchange interaction (J) and transition temperatures are estimated within the 
Ising model of spin Hamiltonians\cite{theory_ising}.
Orbital resolved understanding of the underlying mechanisms have been developed based on Wannier functions\cite{wan-1937},
which are spatially localized linear combination of Kohn-Sham eigen states, and
are known to unambiguously divide the charge density into bonding and atomic orbitals.

\begin{figure}[t]
\centering
\includegraphics[scale=0.17]{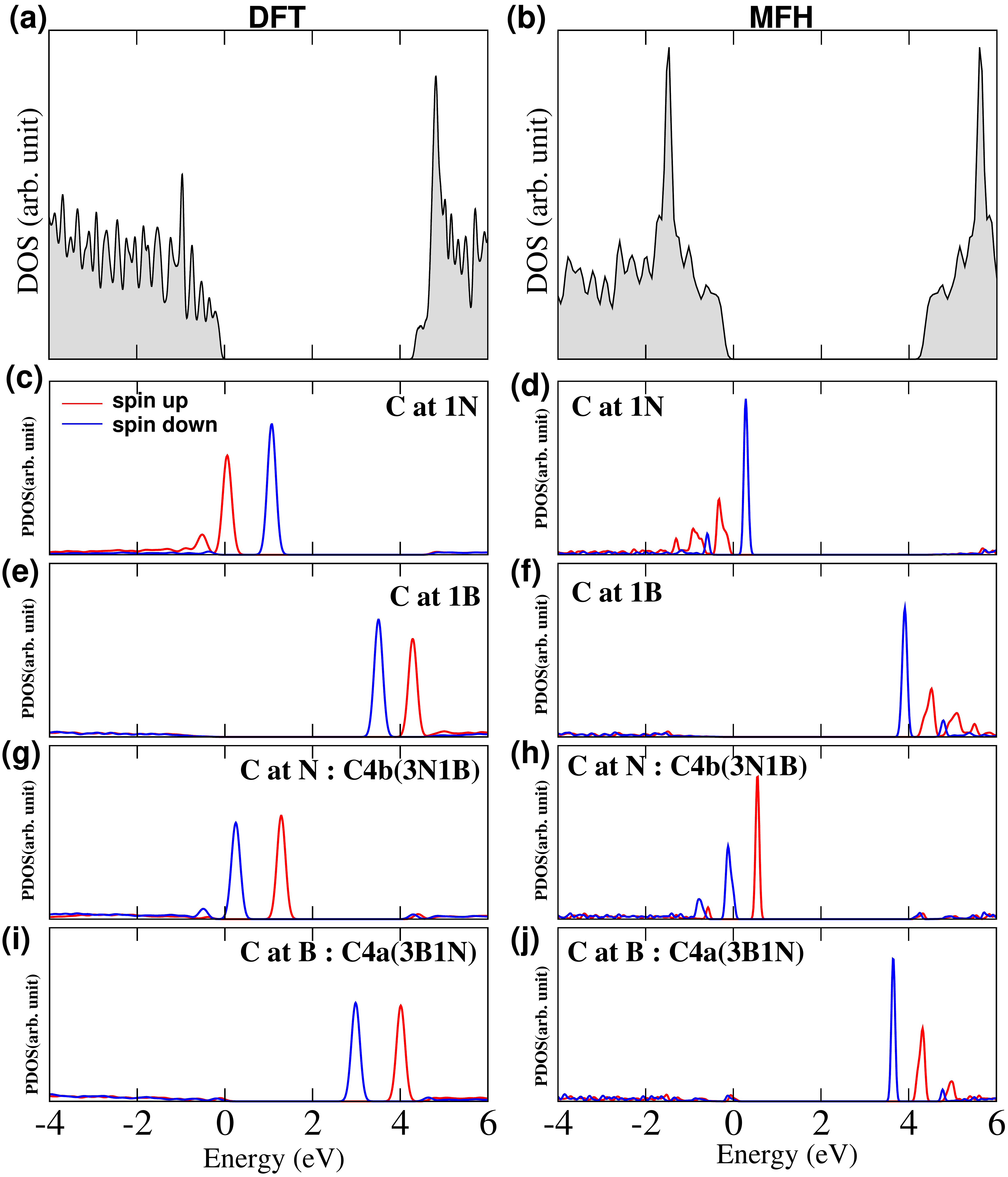}
\caption{Density of states(DOS) of pristine hBN from (a)DFT and (b)MFH model. 
Density of states projected on 2$p_z$ orbital of C-atom substituted at (c,d) single N site, (e,f) single B site, 
(g,h) 3 N and 1 B sites, (i,j) 3 B and 1 N sites, from spin polarized DFT(c,e,g,f) and MFH model(d,f,h,j).}
\label{fig1}
\end{figure}
\section{Computational details}\label{tbparameter}
Equilibrium configurations and spin polarized energetics are calculated within the framework of density functional theory (DFT)
in plane wave basis\cite{QE}, 
using $ultrasoft$ pseudopotentials\cite{usp} and gradient corrected Perdew-Burke-Ernzerhof(PBE) exchange-correlation\cite{pbe} 
functional. PBE results have been refined using the hybrid HSE\cite{QE,hse} approximation of exchange-correlation for better representation
of exchange-interaction.
However, given the computationally intensive nature of HSE calculations, 
only a representative variety of calculations have been refined using HSE. 
Total energies minimized using the BFGS\cite{bfgs} scheme, are converged with 
plane-wave cutoff over 800 eV, k-mesh equivalent to 30$\times$30 for a hBN primitive cell
and forces per atom less than $10^{-4}$ Rydberg/Bohr. 
Energetics of the ground states have been estimated starting from parallel(FM) and anti-parallel(AFM) 
initial alignment of net magnetic moments of neighboring islands.
Within each Gr-island we have considered anti-parallel alignment of spin at neighboring C sites.
In all the calculations reported in this work FM or AFM ordering of the initial condition is retained in the ground state, unless they are non-magnetic.
Therefore in cases where (E$_{FM}$-E$_{AFM}$) is negligible, the corresponding ground states are either nonmagnetic, or are nearly degenerate.
Spin polarized Wannier functions are constructed using subroutines\cite{wnf-pr} interfaced with Abinit\cite{abinit} by maximally aligning the 
Kohn-Sham states with a template of one 2$p_z$ atomic orbital on each C atoms for one of the spins, two 2$p_z$ orbitals on each N atoms
for the two spins, and two $\sigma$-bonding orbitals between all nearest neighbors for the two spins.  

To rationalize a phenomenological model to explain the observed DFT results, 
electronic structure of 2$p_z$ electrons has been also calculated using MFH model, wherein 
\begin{equation}
H_{\small{MF}} =\sum_{i,\sigma}\varepsilon_i c^\dag_{i\sigma} c_{i\sigma} + \sum_{<i,j>,\sigma}{t_{ij} c^\dagger_{i\sigma}} c_{j\sigma} 
 + \sum_{i,\sigma}U_i n_{i\sigma}\langle n_{i\sigma^\prime}\rangle.
\end{equation}
Where $\varepsilon_i$ are the on-site energies($\varepsilon_B, \varepsilon_N, \varepsilon_C$), $t_{ij}$ are nearest-neighbor hopping parameters
$(t_{CC}$, $t_{BN}$, $t_{CN}$, $t_{CB})$ and $U_B$, $U_N$, $U_C$ are the strength of on-site Coulomb repulsion. 
Since the valence shell of B, C, and N are of same principal quantum number($n$=2), we have set $U$ for all sites to that
of C in graphene\cite{rev_mag_gnr}. We have also set the hopping parameters $t_{BN}$, $t_{CN}$, $t_{CB}$ to same as $t_{CC}$ of graphene\cite{rev_mag_gnr}. 
Indeed various literatures\cite{tbH-1,tbH-2,tbH-3} suggest a variation of $U$ and $t$ for different planer three coordinated self-assemblies
with B, C and N to be within 20\% of those used for graphene on the average.
Rather, as evident from Fig.\ref{fig1}(a,b), we have matched the bulk band gap of hBN calculated from MFH to that from DFT 
to tune the on-site energies of B and N, which are the only parameters to vary substantially from that of C. 
After setting the on-site energy for C to zero, we find the on-site energies for B and N to be 3.1 eV and -3.1 eV respectively. 
Comparison of density of states(PDOS) projected on 2$p_z$ orbitals[Fig.\ref{fig1}(c-j)] imply satisfactory qualitative 
agreement between their DFT and MFH descriptions.
The quality of agreement can be incrementally improved quantitatively further by marginally tuning $U$ and $t$ values, although inexhaustibly.
However, our goal to demonstrate the validity of the mechanisms suggested by DFT results
within the MFH model, has been satisfactorily accomplished with the MFH parameters used in this work.

Using first principles data we estimate the exchange coupling parameter $J$ and the corresponding transition temperature(T$_C$) 
by using the Ising model of honeycomb lattice considering only nearest-neighbor coupling,\cite{theory_ising} as,
\begin{equation}
T_C = \frac{2J}{ln(2+\sqrt{3})} ,\label{tc}
\end{equation}
where $-6J=(E_{FM} - E_{AFM})$, and $E_{FM}, E_{AFM}$ are the energies corresponding to FM and AFM configurations obtained from DFT.
We thus refer ($E_{FM}- E_{AFM}$) as the strength of magnetic ordering.
\section{Results and discussions}   
Although there has been a steady rise in the number of computational\cite{afm-rt,c-island}
as well as experimental\cite{ebeam,expt-chbn,expt-fm_BN_GR} 
studies of graphene islands in hBN, attempts to understand the mechanism of mediation of magnetic order by hBN has been limited.
From first principles calculations, magnetic islands in proximity in hBN have been argued to favor AFM ordering as a
means to allow delocalization of spin densities\cite{afm-rt} in conformity with Pauli exclusion principle. 
Similar AFM ordering has been reported in general for isolated C sites in hBN as well\cite{expt-fm_BN_GR,c-dope-2018}, 
although, at some specific proximities and site coverages a less stable FM ordering has also been suggested\cite{expt-fm_BN_GR}
The strength of magnetic ordering has been shown to drastically reduce with increasing separation\cite{afm-rt,c-island}. 
Energetically, FM and AFM ordering of magnetic islands in the ground state has been suggested to be metastable\cite{c-island}
and proposed to be determined by the B rich or N rich nature of the Gr-hBN interfaces around the islands, implying the magnetic
ordering of Gr-islands to be a short-ranged.
On the other hand, flat band based long-ranged mechanisms has been also anticipated to be responsible for FM ordering in C doped 
 hBN\cite{expt-fm_BN_GR}. 
FM ordering of free standing triangular Gr-islands through odd membered C chain has been shown possible 
due to the nearest neighbor AFM ordering of the bridge sites\cite{trignf-chain,grflakes-fm} between Gr-islands,
although such a mechanism is unlikely to be effective in case of triangular islands embedded in hBN.
Attempts so far to explain the observed nature of magnetism in these Gr-hBN hybrid systems have thus remained largely 
speculative, and did not involve the possibility of hBN to take any active role, 
despite the spatial range of magnetic ordering being more than twice the magnitude of primitive lattice constant of hBN. 
We probed this particular aspect in details and indeed found that interactions at the intervening B and N sites to play the
central role in determining the nature of magnetic ordering of the Gr-islands.  

As a source of magnetic moment in hBN, we started with isolated substitution by C denoted as C1(C1a:C@1B, C1b:C@1N),
and have incrementally considered bigger islands made of three coordinated C: 
C4a(C4b) covering three B(N) and one N(B) site, C9 covering six B and three N sites, and C13 covering seven B and six N sites, 
with magnetic moments 2$\mu_B$, 3$\mu_B$ and 1$\mu_B$ respectively\cite{lieb}.
For bigger islands we have considered higher coverage of B site than that of N site, 
since substitution by C is known\cite{formation} to be energetically more favorable at B site than that at N site.

\begin{figure}[t]
\centering
\includegraphics[scale=0.2]{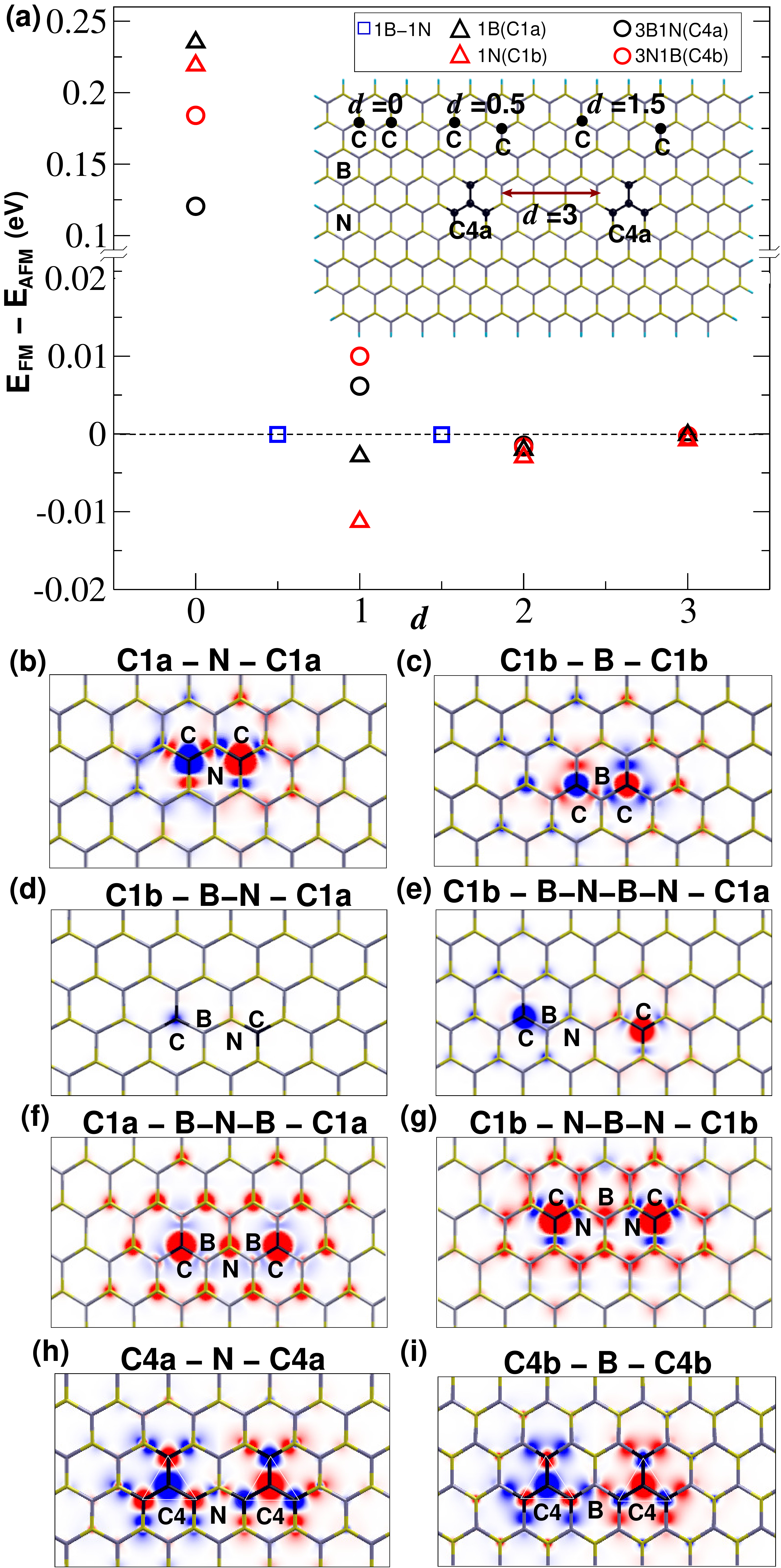}
\caption{ (a) Energy difference(E$_{FM}$ - E$_{AFM}$) for two C1 sites(single substitution) and C4 islands 
for different island separation $d$ within a hydrogen passivated hBN segment.
Spin densities with separation (b,c) $d$=0, (d) $d$=0.5, (e) $d$=1.5, (f,g) $d$=1 between two C1 sites,
and (h,i) for separation $d$=1 between two C4 islands.
All contour plots except (d) are in the range -0.001 to +0.001. (d) is in the range -0.0001 to +0.0001. }
\label{fig2}
\end{figure}
\begin{figure*}[t]
\centering
\includegraphics[scale=0.2]{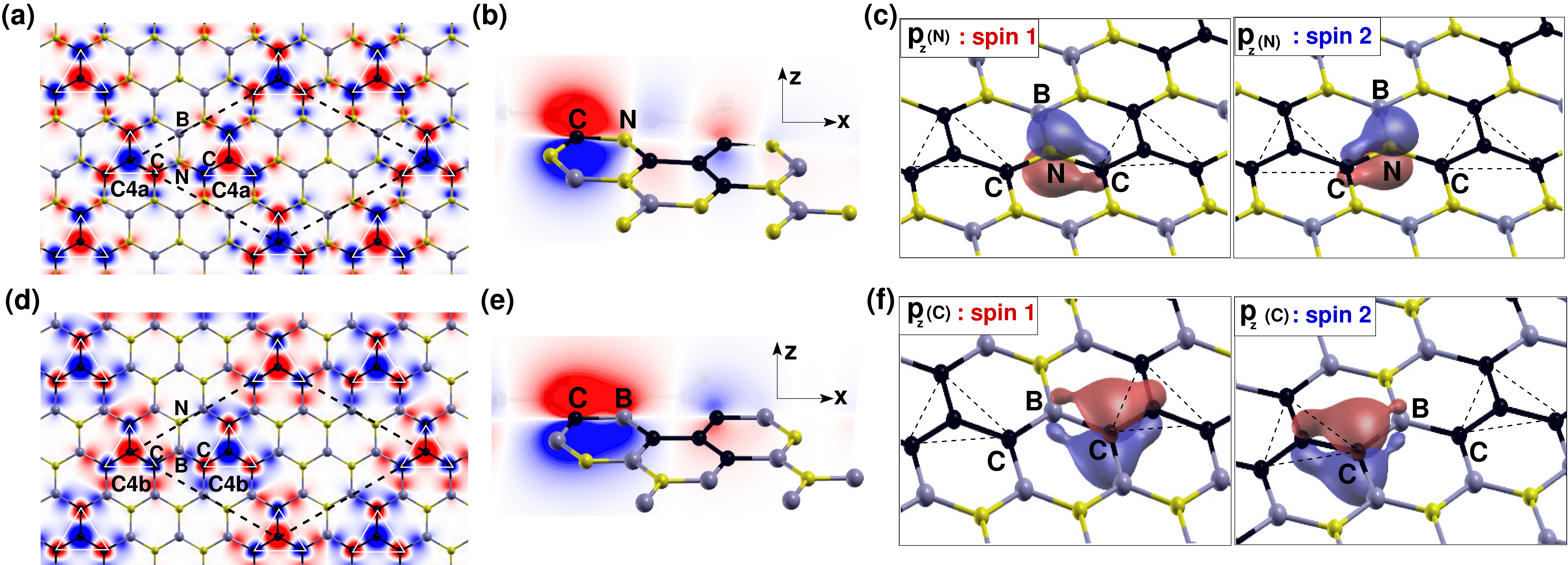}
\caption{Spin density of honeycomb super-lattice with (a) C4a(3B1N), (d) C4b(3N1B)  with $d=0$. 
Planar projection of Wannier functions representing 2$p_z$ orbital of C along (b) C-N and (e) C-B bonds 
for honeycomb super-lattices made of C4a and C4b islands respectively. 
Spin-resolved Wannier function of 2$p_z$ orbital of (c) the bridging N atoms between C4a(3B1N) islands and (f) the outer C atoms
of a C4b(3N1B) island.}
\label{fig3}
\end{figure*}
\subsection{Isolated pair of magnetic islands in hBN segment}
\label{isolated}
In Fig.\ref{fig2}(a) we plot the strength of magnetic ordering as a function of proximity($d$) of two graphene islands. 
As depicted in the inset of  Fig.\ref{fig2}(a), integer(fractional) values of $d$ imply location of the magnetic moments 
in the same(different) sublattice(s), leading to unequal(equal) number B and N sites between the local moments.
As implied in Fig.\ref{fig2}(a), for both C1 and C4 islands, our results suggest
AFM ordering at close proximity ($d$=0) irrespective of whether the islands are rich in B-site(C1a, C4a) or N-site(C1b, C4b). 
For $d$=0, spin densities of the AFM ordered ground state depicted in  Fig.\ref{fig2}(b,c) and (h,i) for C1 and C4 respectively, 
both reveal spatial separation of electrons of opposite spins on the intermediate B and N sites,
hinting at the possibility of a super-exchange\cite{fazekas} like mechanism leading to AFM ordering.
At $d$=1, AFM order sustains between C4 islands but weak FM order emerges between C1 islands.
Such switching of order from strong AFM at $d$=0 to weaker FM at $d$=1 for C1 hints at a competitive scenario of interactions. 
To rationalize the FM order, we note that in the absence of any mechanism to support AFM order,
FM order between the C atoms at same sublattice would have been natural, owing to the on-site Coulomb repulsion driven Mott like
inter-sublattice spin separation induced by the localized 2$p_z$ electrons of C1, as implied in Fig.\ref{fig2}(f,g).
Such a mechanism would favor FM(AFM) ordering of local moments in the same(different) sub-lattice(s). 
The degree of localization of 2$p_z$ electrons is stronger in C1 than in Gr-islands
due to the scope of delocalization of those electrons  in the latter on account of $\pi$ conjugation.
Therefore the Mott like spin separation would be induced more strongly by the 2$p_z$ orbital of C1, 
than by those in C4 or larger islands, which is likely the reason why at $d$=1 the FM order is observed 
only between C1 in the same sublattice, and not between C4.
At $d$=0, such FM order is thus completely suppressed by the AFM order even for C1.

At $d$=0.5 and 1.5 both the mechanisms should have led to AFM order, instead, 
vanishing strength of magnetic ordering is observed for $d$=0.5 and 1.5.
For $d$=0.5, the ground state has negligible magnetic ordering, while for $d$=1.5, the FM and AFM ordered states are effectively degenerate in energy,
as apparent from the spin densities of the corresponding AFM ordered ground state plotted in Fig.\ref{fig2}(d,e).
Thus the AFM and FM order both appear to weaken if the connecting -B-N- pathways allow an equal number of B and N atoms.
The likely reason for the weakening of the AFM is the asymmetric neighborhood of each of the intervening sites along the pathway 
connecting the two moments. For example, in -C-B-N-C- pathway where the C atoms provide the local moments, 
neither B nor N has a symmetric neighborhood. Such denaturing of the super-exchange bridge is evident in Fig.\ref{fig2}(e).
Whereas, in case of a -C-B-N-B-C- (or -C-N-B-N-C-) pathway, the N (or B) atom in the middle 
has symmetric neighborhood which forces formation of a super-exchange bridge on it, leading to AFM ordering between the C atoms.
Weakening of the FM order with equal coverage of the two sublattices by the intervening sites can be understood by noting that 
in general Mott like separation of spins on equal footing among the two sublattices will be hindered due to their different on-site energies.  
On the other hand, with unequal coverage of the two sublattices by the intervening sites the separation of spins need not be on equal footing 
among the two sublattices as evident in Fig.\ref{fig2}(f,g), which conforms with their different on-site energies.
  
C2 islands due substitution at nearest neighboring sites, constitutes C=C dimers with weak localization of 2$p_z$ electron 
on the C atom with N neighbors owing to the effective heteropolarity of the C=C $\pi$ bond. 
C3 islands will be magnetic due to unequal coverage of the two sublattices. 
C2 and C3 are expected to have similar magnetic ordering as C4, owing to weaker or similar localization of 2$p_z$ electrons compared to that in C4.
Notably, at $d$=2, both C1 and C4 islands are weakly FM ordered, implying weakening of the AFM ordering. 
Thus the AFM ordering is stronger than FM ordering in close proximity but has a relatively shorter range than the FM ordering.
Therefore, both the competing mechanisms are driven by the on-site Coulomb repulsion, but originate at different sites, which differ
in occupancy, since AFM (FM) ordering can be generalized to originates at sites occupied by even(odd) number of electrons.

\begin{figure}[b]
\centering
\includegraphics[scale=0.18]{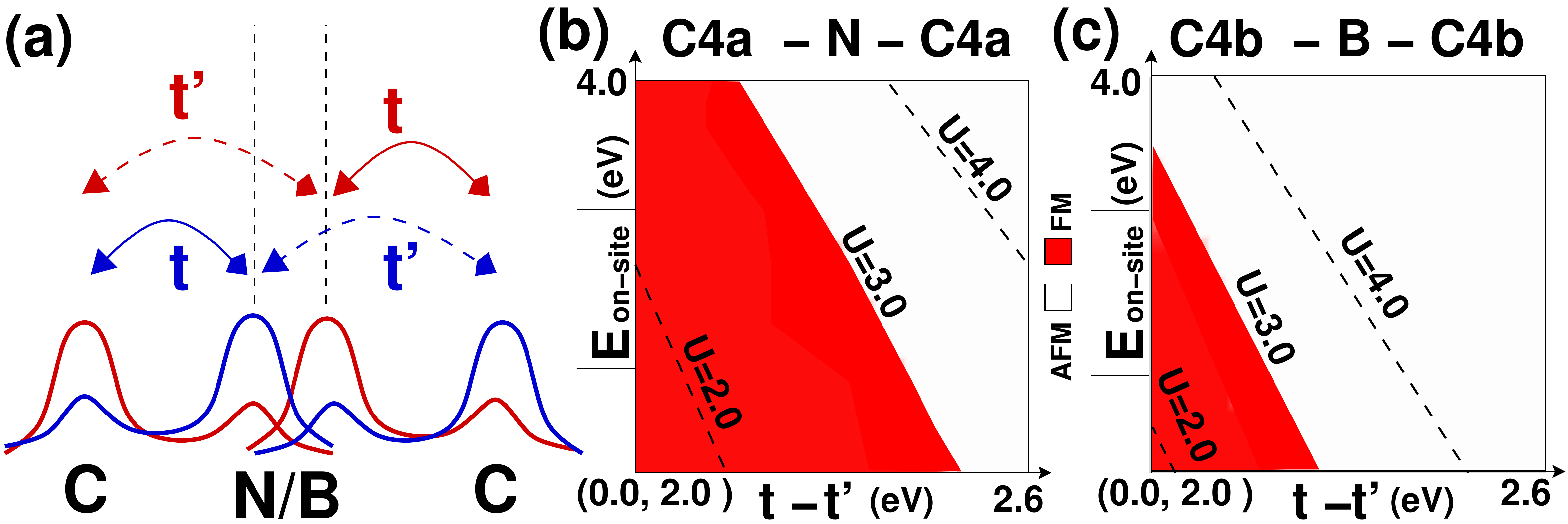}
\caption{(a): Schematic representation of spin dependent hopping mechanism. Phase diagram in terms of total magnetization of the ground state 
calculated as function of (t-t') and on-site energy E$_{on-site}$ for a pair of (b): C4a(3B1N) and (c): C4b(3N1B) islands with $d=0$
 in a finite hBN segment at three different U values(eV). 
For U=4.0 eV and U=2.0 eV the FM-AFM crossover is shown by black dotted line.}
\label{fig4}
\end{figure}

\subsection{Honeycomb lattice of magnetic islands}
In agreement with previous reports\cite{afm-rt}, we also find AFM ordering of sub-nanometer length-scale   
between neighboring Gr-islands in honeycomb lattice.
Effective J estimated using HSE approximation of exchange-correlation confirms that -B-N- zigzag  
connectivity between neighboring Gr-islands favor their AFM ordering, 
more than that due to -B-N- armchair connectivity \cite{afm-rt}.
This is in agreement with the result discussed in the previous section that magnetic ordering between two local moments
in hBN would weakens if they are connected through -B-N- pathway having an equal number of B and N sites, as is the case with armchair connectivity. 
Strength of AFM order, as measured by J between the Gr-islands, is found more if mediated by B than by N, and reduces rapidly beyond $d=3$
for honeycomb lattice.

Spin densities in the vicinity of the intermediate B(N) atom between two C4b(C4a) islands suggest[Fig.\ref{fig3}(a,d)] 
similar spatial separation of electrons of two spins on opposite sides of the B(N) atoms, as seen in Fig.\ref{fig2}(b-e).
To trace the origin of this spin separation we looked at the spatially localized Wannier functions\cite{wnf-pr}    
representing the 2$p_z$ orbitals of C and N.
As evident in Fig.\ref{fig3}(c), the two unpaired 2$p_z$ orbitals of N with opposite spins extend spatially
in opposite directions, resulting into back transfers of opposite spins to the C atoms on its two sides. 
Such a spatial splitting of lone pair is a hallmark of super-exchange pathway.
Similarly on B atom between two C4b islands[Fig.\ref{fig3}(f)], 
the back transfered electrons from the 2$p_z$ orbitals of the two neighboring C atoms are of opposite spins, 
implying spin separation about B as observed in the spin densities[Fig.\ref{fig3}(a,d)]. 
Therefore, on N sites the orbitals offering the super-exchange bridge are fully occupied, while on the B site they
are partially occupied by the back transferred electrons from neighboring N or C sites.
Furthermore, Fig.\ref{fig3}(b,e) imply that the back transfer from C to B is more than 
that to N, which is consistent with the fact that the effective J is more for B mediated AFM than N mediated AFM.

\begin{figure}[t]
\centering
\includegraphics[scale=0.13]{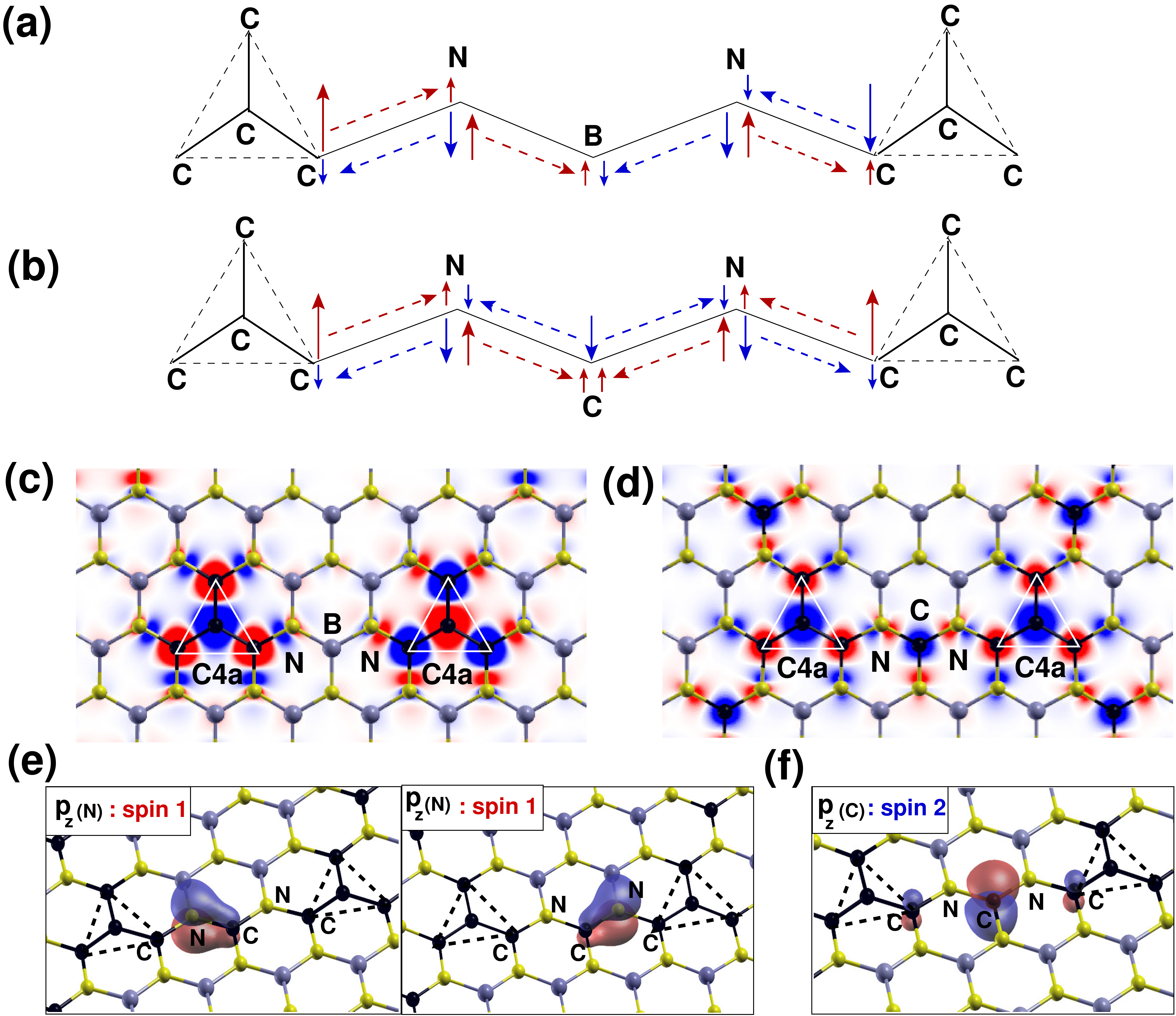}
\caption{Schematic model of (a): AFM ordered C4a(3B1N) islands.
(b): FM ordered C4a(3B1N) islands due to the extra local moment(C) at B site between the C4a islands.
In (a,b) the dashed arrows represent back transfer of electrons to neighboring sites. 
Spin density plot of honeycomb super-lattice of C4a islands with $d=1$ in (c): absence and (d): presence of C 
in the intermediate B sites.
Wannier function representing the 2$p_z$ orbitals of (e): the two N atoms back transferring to the C
in the middle, and (f): of the C atom itself, in the -N-C-N- zigzag pathway between C4a islands.}
\label{fig5}
\end{figure}

\begin{figure}[b]
\centering
\includegraphics[scale=0.11]{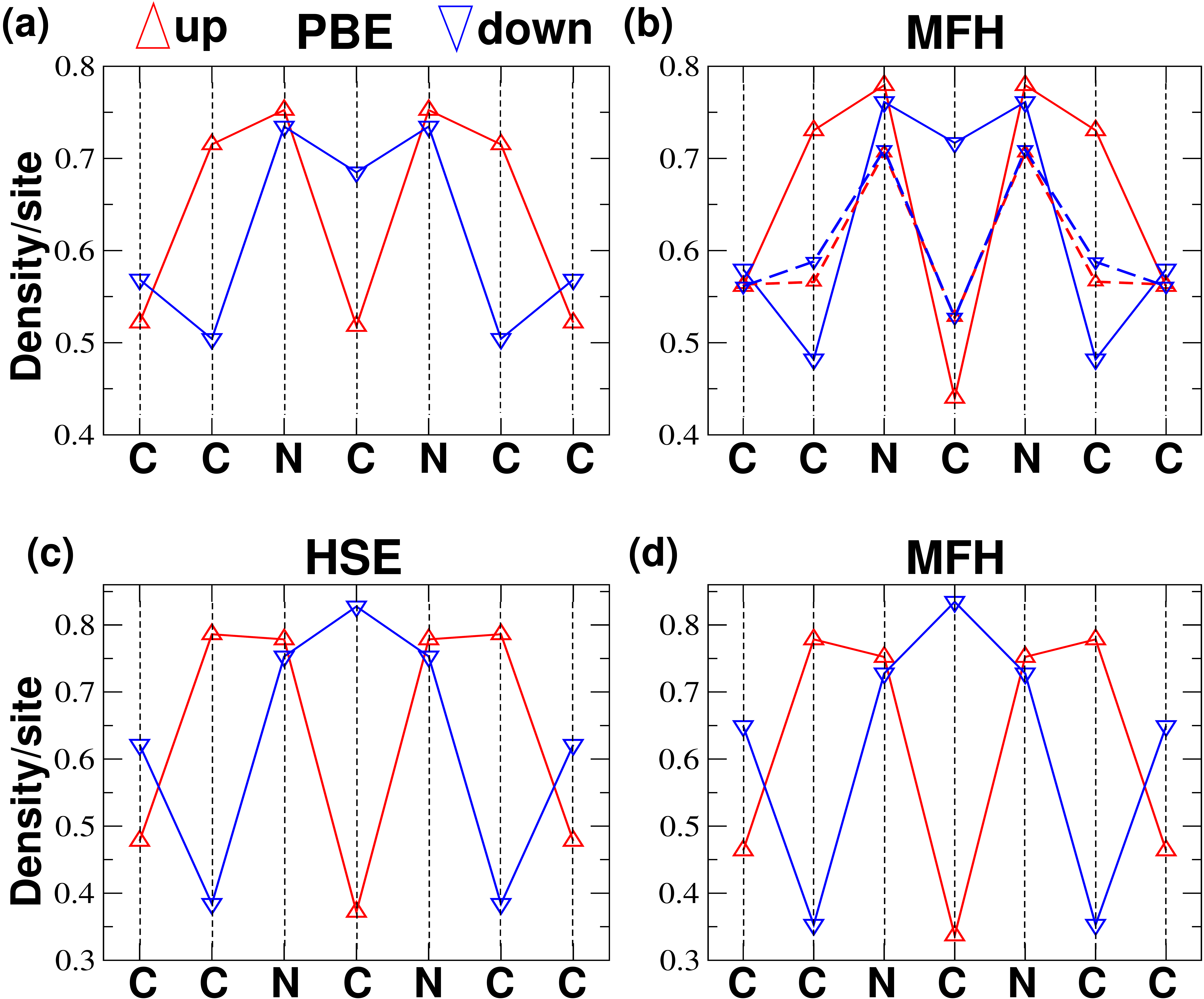}
\caption{Occupancy of 2$p_z$ orbitals of the two spins along zigzag path due to -N-C-N- connectivity 
between C4a(3B1N) islands calculated using (a): PBE and (c)HSE, as well as from MFH with (b): $U$=3.0 eV, and (d): $U$=6.0 eV.
$\Delta$t set to 0.5 eV for (b) and (d).
The occupation represented by the dotted lines in (b) corresponds to $\Delta$t=0.0 eV.}
\label{fig6}
\end{figure}

\begin{figure*}[t]
\centering
\includegraphics[scale=0.2]{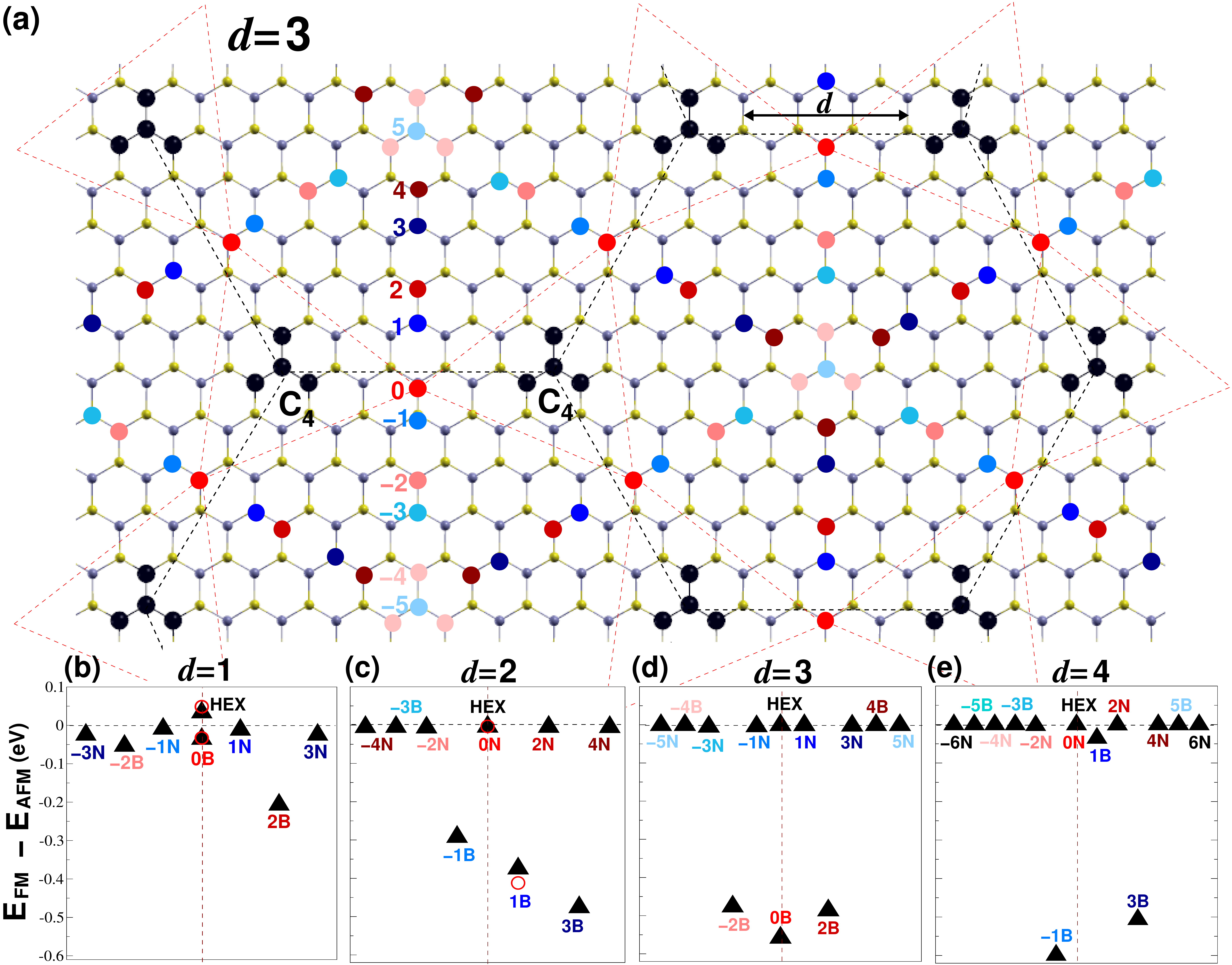}
\caption{(a):Representation of inter-penetrating C4a-X honeycomb-Kagome super-lattices for various intermediated substitution by C at X (B or N) sites
for $d=3$. Energy difference ($E_{FM} - E_{AFM}$) for different X different separations of C4a islands,(b): $d$=1, (c): $d$=2,
(d): $d$=3 and (e): $d$=4. Red circles in (b, c) represents corresponding energy difference with an additional layer of hBN at A-B stacking
beneath the hybrid layer as reported for hBN bilayer. $E_{FM} - E_{AFM} < 0$ implies FM order in honeycomb super-lattice. }
\label{fig7}
\end{figure*}

\subsection{Mechanisms of mediation of magnetic ordering through hBN}
Super-exchange pathway leading to spin selective back transfer in effect indicates 
spin dependent hopping of electrons as a possible realization within a TB framework.
We resort to MFH model to test the relevance of the super-exchange bridge represented by the spin dependent hopping, 
in determining the correct magnetic ordering of the ground state.
Spatial separation of spins suggest symmetric opposite displacement of orbitals of opposite spins away from the host atom, as depicted
in the schematic model[Fig.\ref{fig4}(a)], which implies similar increase in the orbital energies (on-site term) for both the spins.
Accordingly, we consider a pair of C4 islands in close proximity ($d$=0) in the middle of a large hBN segment whose edges are 
sufficiently away from the C4 islands, and calculate AFM, FM and non-magnetic(NM) ground states as a function of spin-asymmetry of hopping
$\Delta t=t_{\uparrow}-t_{\downarrow}$ applied to all the (B/N)-C bonds around the C4 islands, and on-site term of the
(B/N) atoms connected to the C4 islands, and look for the true magnetic ordering of the ground state by comparing  total energies
calculated in FM, AFM and NM conditions. 
We chose an isolated hBN segment in order to avoid the dependence of the magnetic ordering on the choice of periodic unit cell. 

As evident from the TB-MFH based phase diagram shown in Fig.\ref{fig4}(b,c)], the ground state with spin independent
hopping and standard parameters, as discussed in sec.\ref{tbparameter}, is FM ordered with total 4$\mu_B$ magnetic moment, 
which is in disagreement to DFT results.
The AFM ordering of the  ground state, emerges only beyond a threshold value of $\Delta t$, 
and the threshold $\Delta t$ itself decreases with decreasing $U$ as well as increase in on-site term.     
Spin dependent hopping is therefore crucial for the ground state to have AFM ordering of Gr-islands in agreement with DFT result. 
The fact that the $\Delta t$ threshold decreases with decreasing $U$,
reiterates the role of $U$ at B and N sites in mediating the AFM order.
In agreement with the fact the B mediated AFM is stronger than N mediated AFM, the onset of AFM order in the ground state
indeed occurs at a lesser threshold for $\Delta t$ in case of B mediated AFM.
Fig.\ref{fig5}(a)], schematically summarizes the mechanism of propagation of AFM order through -B-N- zigzag pathway
between Gr-islands, where the spatially separated spin polarized 2$p_z$ orbitals at N and B sites act as super-exchange bridges 
between local moments. The dashed arrows in Fig.\ref{fig5}(a,b) indicates the direction of back transfer.

This mechanism thus implies a generic refinement of TB model for bipartite lattices if the sublattices have
different on-site energies, and thereby are occupied differently. 
In fact, the mechanism points to a general property, that the magnetic ordering mediated by a bipartite system
between two local magnetic moments would depend on the degree of asymmetry of the two constituent sublattices and
the degree of  localization the magnetic moment.
With lesser asymmetry of the intervening sublattices, or stronger localization of the moment, 
Mott like spin separation will more likely to lead to FM order between magnetic moments in the same sublattice, 
as we see in graphene with local sublattice asymmetry due to functionalization.
With increasing asymmetry of the intervening sublattices, or weaker localization of the moment,
activation of super-exchange bridge at the sites of the sublattice with higher occupation is more likely to lead
to AFM order of magnetic moments in the same sublattice. 
Since consolidation of both FM and AFM order requires the intervening pathway to be symmetric about the middle, magnetic ordering 
will be weaker in general between local moments at different sublattices. 

\subsubsection{Ferromagnetic ordering}
The mechanism[Fig.\ref{fig5}(a)] also readily suggests that the AFM order can be switched to FM 
if an unpaired electron is present in the -B-N- pathway connecting the two islands, 
as seen on the C site between the N sites in the -B-N- zigzag pathway between two C4a islands in Fig.\ref{fig5}(b). 
Such an unpaired electron can arise from a single substitution by C, or
from another magnetic Gr island located on or close to the intervening -B-N- pathway between the neighboring Gr-islands of the 
honeycomb lattice.
Spin density in Fig.\ref{fig5}(d), compared to that in Fig.\ref{fig5}(c) without the intermediate C between the two C4 islands,
indeed clearly confirms the anticipation of FM order of Gr-islands(C4a) connected by N-C-N- zigzag pathway.
Facilitation of FM order due to spin separation of the N lone pairs is evident in Fig.\ref{fig5}(e),
where the two N atoms on two sides of C is seen to back transfer electrons of same spin (spin1) to C due to 
presence of a 2$p_z$ electron of spin2 at C[Fig.\ref{fig5}(f)], which appears to open a half-metallic bridge connecting the two Gr-islands.
Note that the  2$p_z$ electron at the intermediate C[Fig.\ref{fig5}(f)] mixes with those of the C4 island through anti-bonding
states.

Since even membered -B-N- pathway is expected to suppress mediation of magnetic order as argued above in Sec.\ref{isolated}, 
the generalization of the -N-C-N- pathway for strong FM ordering is -(2d+1 B and N)-C-(2d$^\prime$+1 B and N)-, where d and d$^\prime$ are
integers.                                                                                               
These results suggest that in general an odd number of magnetic Gr-islands
if located within nanometers of each other in hBN, can be FeM ordered amounting to a net magnetic moment.

As evident from Fig.\ref{fig6}(a,c) the FM ordering in the honeycomb lattice and the FeM ordering between honeycomb and kagome lattice 
enhances with HSE exchange-correlation functional compared to that with PBE.
Along the -N-C-N- pathway we have matched the PBE spin density with that due to MFH with U=3.0 eV and $\Delta t$ = 0.5 eV.
Notably, with $\Delta t$ = 0 the FM ordering does not arise and spin separations are minimal[dashed line in Fig.\ref{fig6}(b)]. 
Since Hubbard Hamiltonian in effect evolves into a spin exchange Hamiltonian at high U,
the HSE spin densities along the -N-C-N pathway [Fig.\ref{fig6}(d)] matches with that due to MFH with U=6.0 eV and 
$\Delta t$ = 0.5 eV.
This indispensability of $\Delta t$ being non-zero confirms the central role of spin dependent hopping in rationalizing
the FM ordering of Gr-islands mediated by -N-C-N- pathway.  

\begin{figure}[t]
\centering
\includegraphics[scale=0.38]{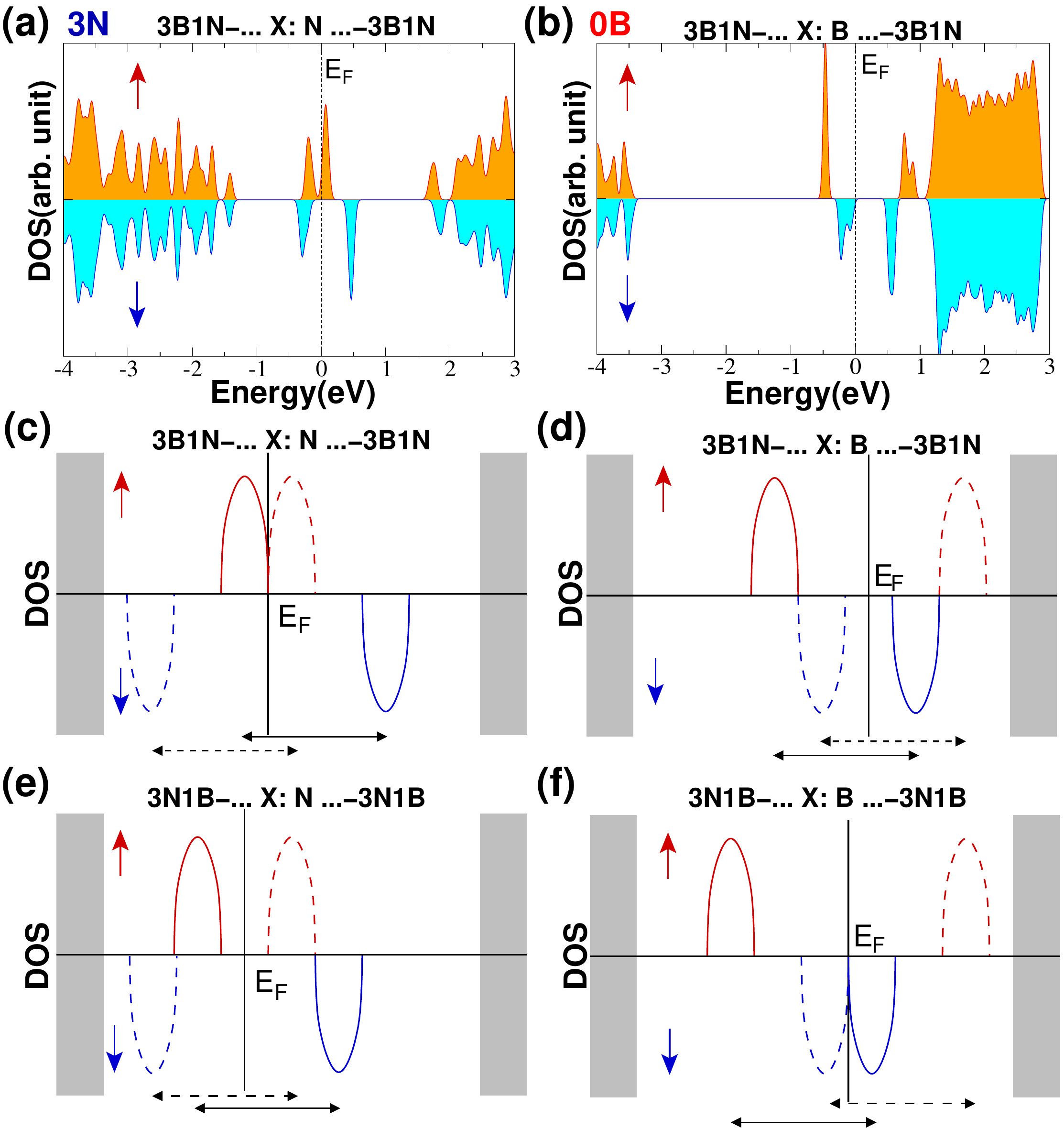}
\caption{Spin-polarized density of states(DOS) for (a): X=3N and (b): X=0B as described in Fig.\ref{fig7}(a,d) for $d$=3. 
(c-f): Schematic presentation of DOS to understand its evolution of half-metallic and FM-Sc phases with different choices of C4 islands 
(C4a or C4b) and X (N or B).}
\label{fig8}
\end{figure}
\subsubsection{Double lattices of Gr-islands in hBN}
As evident in Fig.\ref{fig7}(a), the Gr-islands(C4), and the sites(X) of the unpaired electron, which can be a B or a N site, describe a system of
two inter-penetrating super-lattices: one being the honeycomb super-lattice made by the C4 islands, while the other one
is a twisted Kagome super-lattice made by X sites.
The -B-C-B- or -N-C-N- mediated FM order of Gr-islands suggest a FeM order between the super-lattices.
We now survey the variation in strength of the FM order in the hexagonal super-lattice as a function of location of the X site,
which determines the twist of the Kagome lattice, as depicted in Fig.\ref{fig7}(a).
The energy difference E$_{FM}$ $\textendash$  E$_{AFM}$ plotted in Fig.\ref{fig7}(b-e) for C4a-X honeycomb-Kagome (H-K) super-lattices,
suggest a systematic emergence of strong FM ordering in the hexagonal super-lattice, if the X site allow an odd membered -B-N-
pathway connecting the Gr-islands. As evident in Fig.\ref{fig7}(b-e), this is possible only with X:B(N) for the honeycomb lattice made of C4a(b).
That even membered -B-N- pathway suppress propagation of magnetic order, is reiterated by 
the observation that strong FM order between C4a islands occurs only for X:B, since such X connects to nearest C4a island 
through odd membered -B-N- pathway. X should also be typically within 7\AA{} from the shortest -B-N- zigzag pathway connecting 
two neighboring C4a islands. 
Similar results exist for C4b islands as well as bigger C9(6B3N) islands in the hexagonal super-lattice.
In fact, these results are valid for a general Cm-Cn H-K super-lattices, where Cm and Cn are two magnetic Gr-islands constituting the two lattices,
and can be chosen to be same or dissimilar.
The corresponding $T_C$ estimated using eqn.\ref{tc} 
indicates existence of FM order at room temperature.
Fig.\ref{fig7}(b-e) also suggests a $d$ dependence of strength of the FM order, which can be understood in terms of
the competition between the inherent AFM order of the Gr-islands in the honeycomb lattice and the FM order induced by the
Kagome lattice of X sites. With increasing $d$ the strength of the AFM order reduces, leading to a peak of the FM order, which reduces with further
increase of $d$ beyond 4, implying in effect a length-scale of nanometer.
The impact of having an additional layer of hBN in A-B stacking geometry at a depth of 3.15\AA as known for
hBN bilayer\cite{bilayer-hbn}, is indeed minimal[Fig.\ref{fig7}(b,c)] in terms of retention of ferromagnetism.
\begin{figure*}[t]
\centering
\includegraphics[scale=0.35]{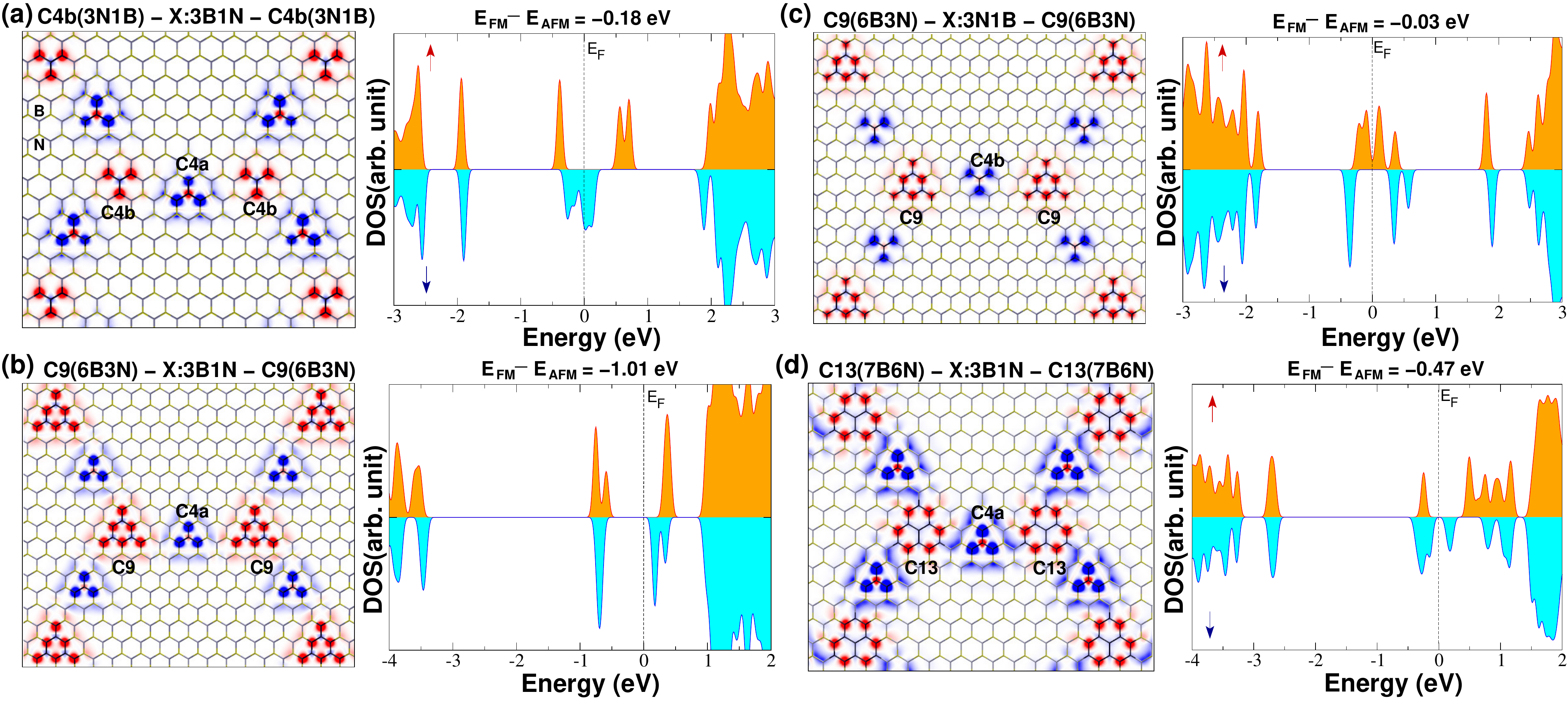}
\caption{Spin density and DOS for (a): C4b-C4a, (b): C9-C4a, (c) C9-C4b, (d) C13-C4a honeycomb-Kagome super-lattices, 
where C9:6B3N,C4a:3B1N, C4b:3N1B and C13:7B6N.}
\label{fig9}
\end{figure*}
\subsection{FM ordered phases}
FM ordering in the honeycomb super-lattice leads to FM-semiconducting(FM-Sc) [Fig.\ref{fig8}(b)] to half-metallic [Fig.\ref{fig8}(a)] phases
depending on X being a B site or an N site in C4a-X H-K super-lattices. Emergence of FM-Sc or half-metallic phases can be understood
in terms of relative shifts in energies of (1): the 2$p_z$ orbitals of the C atoms at the edges of the Gr-islands, and 
(2): those of the C atoms at X site. DOS of the two sets of orbitals are represented by solid and dashed lines respectively 
in the schematic DOS shown in Fig.\ref{fig8}(c,d,e,f), drawn on the basis of orbital projected density of states of different 
configurations shown in Fig.\ref{fig7}.
The relative shift in energies of these two sets of orbitals, which are mechanistically of opposite spins,
can be understood in terms of the difference in their localization owing to their different size(C4 and C1) 
and distribution(honeycomb and Kagome).
Fig.\ref{fig8}(e,f) and Fig.\ref{fig8}(c,d) suggests that the  properties of the C4b-X:B(N) and C4a-X:N(B) 
H-K super-lattices will be similar. 
With same C4a islands in the honeycomb lattice, Fig.\ref{fig8}(c) and Fig.\ref{fig8}(d) explains emergence of half-metallic and FM-Sc phases due to X:N and X:B respectively,  in agreement with DFT results shown in Fig.\ref{fig8}(a) and Fig.\ref{fig8}(b).
Fig.\ref{fig8}(c) and Fig.\ref{fig8}(f) suggests emergence of half-metallic windows of opposite spins at Fermi energies 
due to C4a-X:N and C4b-X:B H-K super-lattices respectively, 
which is consistent with the DOS plotted in Fig.\ref{fig9}(c) and (a) respectively, where the Kagome lattices (X) are made of N site rich C4b and B site rich C4a islands instead of a single C at N or B site respectively. 
The robustness of the half-metallic phases with X:C4a/C4b suggests that the key to make the half metallic window broader 
is to increase delocalization of the  spin polarized electrons at the Kagome lattice.

Fig.\ref{fig8}(d) and (e) suggests FM-Sc phases in C4a-X:B and C4b-X:N H-K super-lattices,
which are consistent with the FM-Sc phases shown in Fig.\ref{fig9}(b) and (d) for C9-C4a and C13-C4a H-K super-lattices. 
Notably, the narrowing of FM-Sc band-gap from  Fig.\ref{fig9}(b) to (d) can possibly be attributed to the 
increased delocalization of electrons in C13 than in C9 due to enhanced $\pi$ conjugation in the former.
The coinciding DOS of the two spins in Fig.\ref{fig9}(b) near Fermi energy can be understood by noting that the degree of localization
of the 2$p_z$ electrons are similar in C4 and C9 islands.
These results thus lead to a simple thumb-rule that a half-metallic(FM-Sc) phase is expected is the Gr-islands in the honeycomb and the 
Kagome lattices are rich in substitution at dissimilar(similar) sites out of B or N, and increased delocalization of 
2$p_z$ electrons in the Gr-islands constituting the honeycomb(Kagome) lattice would lead to wider(narrower) half-metallic(FM-Sc) window(gap).
\section{Conclusion}
To conclude, we show that  Coulomb repulsion driven spatial separation of lone pairs and back transferred electrons 
of opposite spins on N and B atoms, implying in effect a super-exchange pathway,
to be the primary mechanism for the anti-ferromagnetic order mediated by -B-N- pathway between neighboring magnetic Gr-islands 
in hexagonal boron nitride. A weaker ferromagnetic order between local moments in the same sublattice also exists 
due to the generic Mott like inter-sublattice spin separation driven by 
on-site Coulomb repulsion, but suppressed by the super-exchange mechanism effective between graphene-islands, owing to
weaker localization of 2$p_z$ orbitals with increasing island size.
The mediated magnetic order becomes ferromagnetic if an odd number of unpaired spin interjects 
the -B-N- pathway, implying in general a finite net magnetic magnetic moment due to odd number of magnetic islands in proximity.
 An inter-penetrating system of honeycomb and twisted Kagome super-lattices of 
magnetic graphene-islands in hBN is proposed, wherein, the two ferri-magnetically ordered super-lattices should 
retain a net non-zero magnetic moment at room temperature, besides constituting a ferromagnetic semiconductor 
or a half-metal depending on the nature of Gr-islands in the two lattices.

\subsection{Acknowledgement}
This work has been performed in a high performance computing facility funded by 
Nanomission(SR/NM/NS-1026/2011) of the Dept. of Sci. and Tech. of the Govt. of India. RM acknowledges
financial support from the Dept. of Atomic Energy of the Govt. of India.


\end{document}